\documentclass[prl,amstex,amsmath,amssymb,superscriptaddress,showpacs,
twocolumn]{revtex4}

\usepackage{graphicx,dcolumn,epsfig}

\begin{document}

\title{Instantaneous-Shape Sampling for Calculation of the 
Electromagnetic Dipole Strength in Transitional Nuclei}

\author{S.~Q.~\,Zhang $^*$}
\affiliation{Institut f\"ur Strahlenphysik,
             Forschungszentrum Dresden-Rossendorf, 01314 Dresden, Germany}
\thanks{On leave from School of Physics and State-Key  Lab. 
Nucl. Phys. and Tech. Peking University, Beijing 100871, PR China}
\author{I.\,Bentley}
\affiliation{Department of Physics, University of Notre Dame,
             Notre Dame, IN 46556, USA}
\author{S.\,Brant}
\affiliation{Department of Physics, Faculty of Science, University of Zagreb, 
10000 Zagreb, Croatia}
\author{F.\,D\"onau}
\affiliation{Institut f\"ur Strahlenphysik,
             Forschungszentrum Dresden-Rossendorf, 01314 Dresden, Germany}
\author{S.\,Frauendorf}
\affiliation{Institut f\"ur Strahlenphysik,
             Forschungszentrum Dresden-Rossendorf, 01314 Dresden, Germany}
\affiliation{Department of Physics, University of Notre Dame,
             Notre Dame, IN 46556, USA}
\author{B.\,K\"ampfer}
\affiliation{Institut f\"ur Strahlenphysik,
             Forschungszentrum Dresden-Rossendorf, 01314 Dresden, Germany}
\author{R.\,Schwengner}
\affiliation{Institut f\"ur Strahlenphysik,
             Forschungszentrum Dresden-Rossendorf, 01314 Dresden, Germany}
\author{A.\,Wagner}
\affiliation{Institut f\"ur Strahlenphysik,
             Forschungszentrum Dresden-Rossendorf, 01314 Dresden, Germany}
\date{\today}

\begin{abstract}
Electromagnetic dipole absorption cross-sections of transitional nuclei
with large-amplitude shape fluctuations are calculated in a microscopic way
by introducing the concept of Instantaneous Shape Sampling.  The concept
bases
on the slow shape dynamics as compared to the fast 
dipole vibrations. The elctromagnetic dipole
strength is calculated by means of RPA for the instantaneous shapes,
the probability of which is obtained by means of IBA. Very good agreement with 
the experimental absorption cross sections  near the nucleon emission 
threshold is obtained.
\end{abstract}

\maketitle

Photo-nuclear processes, as the absorption of a photon inducing the 
emission of a neutron or the emission of a photon after neutron 
absorption, are key elements in various astrophysical scenarios, 
like  supernovae explosions or
$\gamma$-ray bursts,
%\cite{XXX} 
as well as in simulations for nuclear technology. 
For a quantitative description of the relevant nuclear reactions
one needs to know  the 
$\gamma$-absorption cross section and the reemission probability, 
being determined by the dipole strength function. 
%The reactions take place in the Gamov window, which is an
%energy interval of a few MeV above the neutron emission threshold 
%( or thresholds for the emission of other particles).
Direct measurements 
of the strength function in the relevant 
energy range (typically 6-10 MeV in medium-heavy nuclei) are not possible 
nowadays for most of the unstable nuclei passed in  violent stellar events. 
Theoretical models that provide reliable prediction of the dipole strength 
function are therefore of utter importance.
Aside from the astrophysical applications understanding 
the mechanisms that determine the structure of the dipole strength function
in this energy region is a challenge of its own to nuclear theory. 
The present Letter 
proposes and tests a new
approach, which we call Instantaneous Shape Sampling (ISS).
It combines the microscopic Random Phase Approximation (RPA) for 
dipole excitations  with  
the phenomenological Interacting Boson Approximation (IBA)
for a dynamical treatment of the nuclear shape. ISS
allows one to calculate the dipole strength function of the 
many  transitional nuclei ranging between the regions of
spherical and well deformed shape.
             
Traditionally one
employs phenomenological expressions for the dipole strength function 
\cite{Ripl2}, which are based on the classical model of a damped collective 
Giant Dipole Resonance (GDR). The photo-absorption cross section
$\sigma_\gamma$ of the GDR is approximated by a Lorentzian 
curve \cite{Axel62,Kadmenskii83,Kope90,Raus00}, 
which may include corrections 
for nuclear deformation  \cite{Raus00}. The damping
width of the Lorentzian is treated as a parameter that is
adjusted to the experiment. However, 
the available data have not yet allowed stringent tests  of this
extrapolation toward the low-energy tail of the GDR \cite{Kope90}.
% The situation is engraved by the fact that 
%different functional forms of $\Gamma(E)$ have been suggested 
%in the literature. 
Therefore microscopic approaches that treat 
at least a substantial part of the damping explicitly promise improved
predictive power.

\begin{figure}[t]
\vspace*{-1.5cm}
\includegraphics[height=12cm]{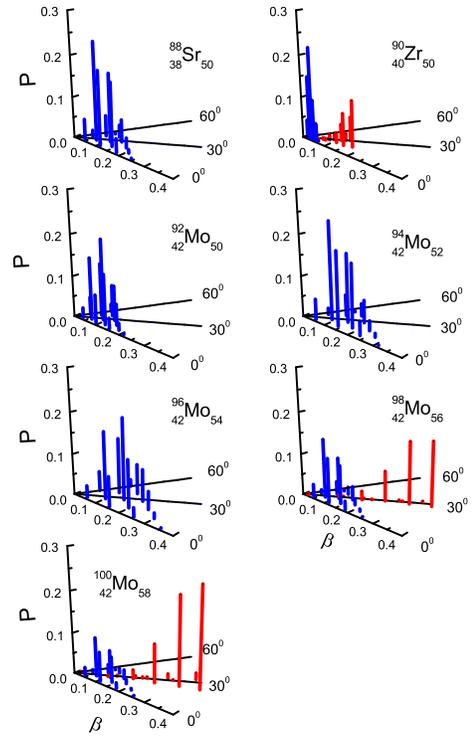}
\vspace*{-1cm}
\caption{(Color online) Probability distributions of 
the instantaneous nuclear shapes
over the $\beta-\gamma$ plane. Families of coexisting shapes are distinguished
by their color.}
\label{distributions}
\end{figure}

\begin{table}
\caption{IBA parameters $~\zeta$, $\chi$, $e_B$, 
and equilibrium deformation parameters $~\beta$, $\gamma$  calculated
by means of the micro-macro method. 
In case of shape coexistence, two sets
are listed. Their respective proportion in the ground state is given
in percentage.}
\begin{tabular}{clclccc}
  \hline\hline
  $^A$X      & $~\zeta$ & $\chi$ & 
$e_B$ & $\%$ & $~\beta$ & $\gamma$ \\ \hline
  $^{88}$Sr &  0.0    & -1.20   & 0.043    &  100&    0.0  & $0^\circ$ \\
  $^{90}$Zr &  0.0    & -1.20   & 0.013    &   64&    0.0  & $0^\circ$ \\
            &  0.60   & -0.31   & 0.040    &   36&         &           \\
  $^{92}$Mo &  0.25   & -1.32   & 0.040    &  100&    0.0  & $0^\circ$ \\
  $^{94}$Mo &  0.29   & -1.20   & 0.064    &  100&   0.02  &$60^\circ$ \\
  $^{96}$Mo &  0.20   & -1.32   & 0.069    &  100&   0.10  &$60^\circ$ \\
  $^{98}$Mo &  0.0    & -1.20   & 0.053    &   60&   0.18  &$37^\circ$ \\
            &  0.59   & -0.03   & 0.106    &   40&         &           \\
  $^{100}$Mo&  0.0    & -1.20   & 0.053    &   40&   0.21  &$32^\circ$ \\
            &  0.61   & -0.10   & 0.106    &   60&         &           \\
  \hline\hline
\vspace*{-1cm}
\end{tabular}
\label{IBAparameters}
\end{table}

RPA \cite{Wambach88}
is the standard microscopic approach to the dipole strength 
function. It takes the Landau fragmentation 
into account, which describes the coupling of the collective dipole
vibration to the  particle-hole excitations. It also describes
the splitting of the GDR caused by a static deformation of the mean
 field. However both effects  account only for a fraction of the observed
width of the GDR. In order to obtain the experimental 
damping width of the GDR one must include, at least, the coupling to  
the two-particle two-hole states  \cite{Wambach88}. One may distinguish 
between two types of such states: the combination of the GDR
with incoherent particle-hole
 excitations and its combination with coherent
collective excitations. The coupling to the 
former is analogue to the collisional damping of Fermi liquids. The latter
has been studied for spherical nuclei
by means of particle-phonon coupling models, such 
as QPM \cite{sol92,QPCT}, QTBA \cite{QTBA}, and QRPA-PC \cite{QRPA-PC}, 
which, however, meet
principle problems in  transitional nuclei. 
  
We suggest an alternative approach. 
Out of the huge space of collective excitations  
coupling to the GDR we select only the low-energy 
collective quadrupole excitations. They represent
the softest mode, which couples most strongly to the dipole mode.
The coupling,  coherently and incoherently, to the other excitations 
 is taken into account 
by averaging with a Lorentzian of the type that arises 
from collisional damping \cite{Kadmenskii83,Wambach88}.
In the following we simply refer 
to this as ``collisional damping'' (CD).  
The typical  energies
of the quadrupole excitations are below 1 MeV, i.e. about a factor of 
10 less than the energy of the dipole excitations.  Since the 
quadrupole motion is ten times slower than the dipole one we use
the adiabatic approximation: By means of RPA, 
we calculate the dipole absorption cross section 
$\sigma_\gamma(E,\beta_n,\gamma_n)$ 
for a set of instantaneous deformation parameters $\{\beta_n,\gamma_n\}$ 
of the mean field. We find
 the probability $P(\beta_n,\gamma_n)$ of each shape being present in the 
ground state by means of IBA-1,
and obtain the total cross section as the incoherent sum 
of the instantaneous ones, 
\begin{equation}\label{adiabat}
\sigma_\gamma(E)=\sum_nP(\beta_n,\gamma_n)\sigma_\gamma(E,\beta_n,\gamma_n).
\end{equation}
In other words, we assume that the quadrupole deformation does not
change during the excitation of the nucleus by the absorbed photons, 
which sample the instantaneous shapes of the nucleus in the 
ground state. Accordingly
 we suggest the acronym ISS-RPA for the approach. 

Ref. \cite{LeTourneux} studied
a collective dipole vibration
coupled to a harmonic quadrupole vibration and showed 
that the adiabatic approximation rather well reproduces the exact
spreading. The ISS  
expression (\ref{adiabat}) is of very general nature. All versions of RPA
based on a deformed mean field could be used to calculate 
$\sigma_\gamma(E,\beta_n,\gamma_n)$. Likewise, any model describing the
the collective quadrupole mode could be used to obtain $P(\beta_n,\gamma_n)$.

We adopt the quasiparticle version of RPA (QRPA) 
described in \cite{Doen07}, 
which combines a triaxial potential with separable interactions. 
We changed to 
a Woods-Saxon potential with ``Universal Parameters''.
%\cite{Kaha89}.
The pairing gaps are adjusted to the even-odd mass differences.
The E1-part of $\sigma_\gamma$ is calculated with  
an isovector dipole-dipole interaction, the strength of which is adjusted to 
the experimental position of the maximum of the GDR. 
The M1-part is calculated with a repulsive isovector spin-spin
 and an isoscalar quadrupole-quadrupole	interaction
(for details cf. \cite{M1}). The $\sigma_\gamma(E)$ are calculated 
using the strength function method with a resolution of 100 keV.

We  describe collective quadrupole mode by means of
IBA-1 \cite{IBA1}, which is known to
well reproduce the development of energies and E2-transition probabilities
from spherical to well deformed nuclei through the transition region.
We use the ``extended consistent Q formalism''. The 
 Hamiltonian and the
E2-transition operators are given in \cite{McCutchan04}. 
The parameters $\zeta,~\chi,~e_B$ of the model 
are listed in Table \ref{IBAparameters}. For practical reasons
we fix the  boson number to $N_B=10$, which turned out to be a sufficient
flexible basis. 
The probability distribution $P(\beta_n,\gamma_n)$ is generated
by means of the method suggested in \cite{Tonev07}. We consider the two
scalar operators 
\begin{eqnarray}
\hat q_2=\left[ Q^{\chi} \otimes Q^{\chi} \right]_{0},\\
\hat q_3=\left[ Q^{\chi} \otimes 
\left[ Q^{\chi} \otimes Q^{\chi} \right]_{2} \right]_{0}, 
\end{eqnarray}
which are formed by angular momentum coupling from the IBA quadrupole
operators $Q^\chi_\mu$. The two commuting operators $\hat q_2$ and
$\hat q_3$ are diagonalized in the space of states generated by
coupling 10 bosons  to
zero angular momentum. The resulting  eigenvalues $q_{2,n}$ and
$q_{3,n}$ and the eigenstates $|n\rangle$ are linked to the
deformation parameters as  follows
% the prescription given in    
% \cite{Werner00} for the ground state {\it expectation values},
\begin{equation}
\beta_n^2=\sqrt{5}\left(\frac{4\pi e_B}{3ZeR^2}\right)^2q_{2,n},~
\cos{3\gamma_n}=\sqrt{\frac{7}{2\sqrt{5}}}\frac{q_{3,n}}{(q_{2,n})^{3/2}}.
\end{equation}  
 The probabilities $P(\beta_n,\gamma_n)=|\langle 0^+_1|n\rangle|^2$
are the projections of the IBA ground state $|0^+_1\rangle$
on the set $|n\rangle$.
The method is easily generalized to the case of shape coexistence. 
 In this case one has two families of states, $a$ and $b$. They are described
by two sets of IBA parameters and the mixing coefficients $c^2_a$ and 
$c^2_b=1-c^2_a$, which give the fraction of each family in the ground state.
We generate  $P(\beta_{na},\gamma_{na})$ 
and  $P(\beta_{nb},\gamma_{nb})$ separately. 
The total distribution is then given by
$c^2_\nu P(\beta_{n\nu},\gamma_{n\nu}),~\nu=a,b$, and the sum in
(\ref{adiabat}) runs over all combinations $\{n,~\nu\}$.

\begin{figure}[t]
\includegraphics[width=9cm]{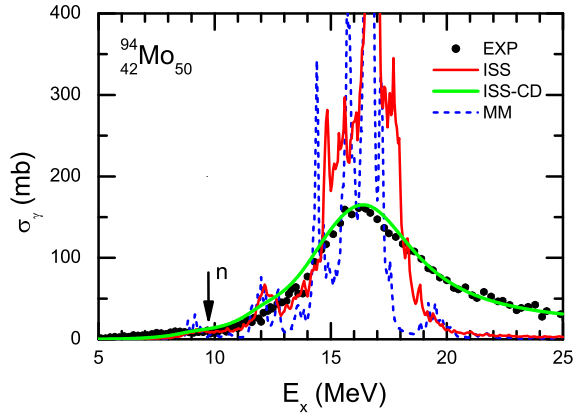} 
\vspace*{-1cm}
\caption{(Color online) Photo-absorption cross section of $^{94}$Mo.  
MM: RPA for the equilibrium deformation, ISS: RPA 
averaged over the probability distributions for shapes
 in Fig.~\ref{distributions}, ISS-CD: ISS 
folded with a Lorentzian of width $\alpha E^2$. Experimental data 
from \cite{Mo-data}}
\label{94Mo}
\end{figure}

We applied the method to the nuclides  
$^{88}$Sr, $^{90}$Zr and $^{92-100}$Mo, for 
which the combination of earlier $(\gamma,n)$ 
measurements \cite{Ripl2} with recent
$(\gamma,\gamma')$ experiments at the ELBE facility 
\cite{Mo-data,98100Mo-data,88Sr-data,90Zr-data} provided
absorption cross sections $\sigma_\gamma(E)$ 
covering the whole energy range from the GDR down to few MeV.  
The IBA parameters
were obtained by fitting the energies and $B(E2)$ values
of the lowest $0^+,~2^+,~4^+$
states taken from the ENSDF data base and from \cite{Garrett03,Cata90}.
We set the boson number $N_B=10$, but otherwise
followed Refs. \cite{Garrett03,Cata90}, where we took the 
mixing compositions from. Fig.~\ref{distributions} demonstrates that
the resulting instantaneous shapes are widely distributed across the
$\beta - \gamma$ plane, which reflects the transitional nature of 
the considered nuclei. In the cases of  $^{90}$Zr and $^{98,100}$Mo the two 
coexisting families of shapes are clearly recognized (blue and red).

\begin{figure}[t]
\vspace*{-.5cm}
\includegraphics[width=7cm]{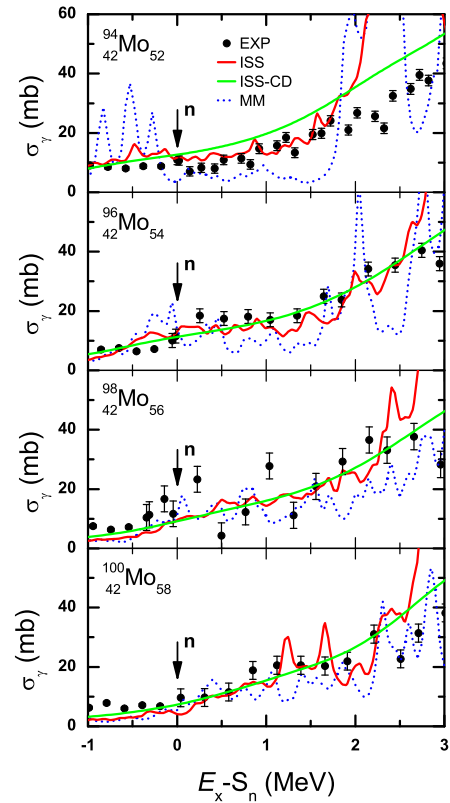} 
\vspace*{-.5cm}
\caption{As Fig.~\ref{94Mo} for the energy range around the neutron emission
threshold, where   
$S_n=$9.68, 9.15, 8.64, 8.29 MeV for $N=52,~54,~56,~58$, respectively.
Data from \cite{Mo-data,98100Mo-data}.}  
\label{Mo}
\end{figure}

\begin{figure}[t]
\vspace*{-.5cm}
\includegraphics[width=7cm]{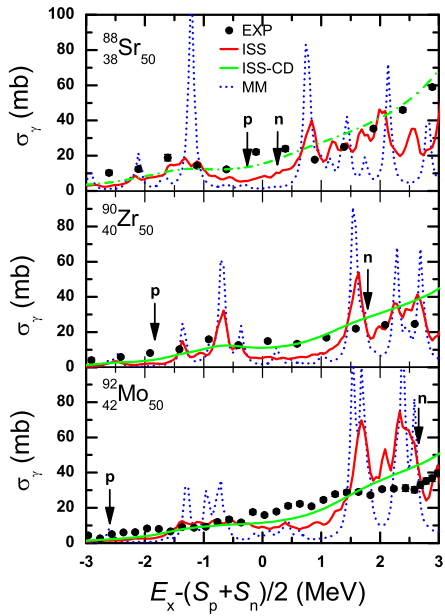} 
\vspace*{-.5cm}
\caption{ As Fig.~\ref{Mo}. The mean of the proton and 
neutron thresholds are
$(S_p+S_n)/2=$10.86, 10.16, 10.06 MeV for $Z=38,~40,~42$ , respectively.
Data from \cite{88Sr-data,90Zr-data} }
\label{N50}
\end{figure}

Figs.~\ref{94Mo}, \ref{Mo}, and \ref{N50} compare the results with the 
experimental data.
We include the RPA results for the
equilibrium deformations (labelled by MM in the figures) 
listed  in Tab. \ref{IBAparameters}, which were 
calculated by means of the micro-macro method
\cite{Doen07,M1}. Let us first consider
 $^{94}_{42}$Mo$_{52}$ shown in Figs.~\ref{94Mo}, and \ref{Mo}. It
 has a spherical equilibrium shape, but  pronounced 
transitional character, which is demonstrated
by the wide distribution of shapes in Fig.~\ref{distributions}.
The spherical RPA (MM) shows the expected  
strong fluctuations of
  $\sigma_\gamma$, reflecting the degeneracy
of the spherical single particle levels. 
The  substantial Landau fragmentation generates dipole strength in the 
threshold region which is of the order of the experimental one. 
Sampling the different instantaneous shapes (ISS) 
largely wipes out
the strong spikes, because the spherical degeneracy is lifted. 
It shifts additional strength into the region below 14 MeV, such that
$\sigma_\gamma$ comes close to the data. At the center of the GDR, the 
ISS peak is broader and lower
than the spherical RPA one. However,
it is still too narrow and too high because the collisional 
damping is missing  \cite{Wambach88}. We include it 
by folding the ISS cross section with a Lorentzian  
of  width $\Gamma=\alpha E^{2}$, 
which depends on the photon energy as expected for collisional
damping \cite{Kadmenskii83,Wambach88}. The coefficient   
$\alpha$ is chosen to reproduce the experimental
$\sigma_\gamma$ at the maximum of the GDR, which gives
$\alpha=$0.0105 MeV$^{-1}$ for the neutron number
$N=$50 and 0.014 MeV$^{-1}$  for $N>50$.
The resulting curve labelled by ISS-CD reproduces the
data very well. The agreement of ISS-CD with the data is as good as  
in Fig.~\ref{94Mo} \
for the other nuclides not shown on this scale.

Figs.~\ref{Mo} and \ref{N50} zoom
into the region near the nucleon emission thresholds.
ISS  well describes the experimental $\sigma_\gamma$ on the average,
while still fluctuating too much. The fluctations are wiped out
in the
ISS-CD curve, which reproduces the data very well. The
inclusion of collisional damping (ISS-CD vs. ISS)
hardly shifts additional strength from the GDR to the
considered energy region.
Hence, the dipole strength near the nucleon emission threshold is 
determined by the Landau fragmentation of the instantaneous shapes,
each of which contributes with its probability being present in 
the ground state. 
%The distribution of the dipole strength
%over the Landau fragments depends sensitively on the position
%and structure of the single particle states in the instantaneous
%deformed potentials, which, thus, are
%essential microscopic ingredients for predicting the dipole strength.  

In order to quantify the contributions to the width of the GDR peak,
we folded the MM and ISS results for $^{94}_{42}$Mo$_{52}$
 with a Lorentzian of constant width
$\Gamma$, chosen such that the peak height agreed with 
the experiment. For the MM case we found $\Gamma$=4.6MeV and for
ISS $\Gamma$=4.0 MeV. Since the experimental value is
$\Gamma_{exp}$=5.7 MeV,
the contribution of the Landau fragmentation is 
\mbox{$\Gamma_{LF}$=(5.7-4.6) MeV= 1.1MeV}. ISS contributes 
\mbox{$\Gamma_{ISS}$=(5.7-1.1-4.0) MeV=0.6 MeV}. The strongest contribution
$\Gamma_{CD}$= 4.0 MeV results from collisional damping.
Hence for the ``spherical'' nucleus $^{94}_{42}$Mo$_{52}$ the distribution of 
dipole strength near the peak of the GDR is
dominated by the collisional damping whereas Landau fragmentation and
shape fluctuations dominate in the low-energy tail of the GDR.
The contribution of shape fluctations to the peak width increases with
$N$, going over to the splitting into two peaks characterizing a well
deformed shape. Applying the same procedure to the other Mo-isotopes,
we found for the collisional
damping width \mbox{$\Gamma_{CD}$=(4.0, 3.6, 3.5, 3.9) MeV} 
and for the combined contribution 
from the fluctuating shape and Landau fragmentation
  \mbox{$\Gamma_{LF+ISS}$=(1.7, 2.7, 2.5, 4.0) MeV}
for $N$=94-100, respectively.

We reproduce  the experimental
$\sigma_\gamma$  near the nucleon emission thresholds   
 using input parameters that are fixed by
{\it independent} information, namely the
location of the GDR and low-lying quadrupole
excitations. This holds the promise that the concept of ISS-RPA
will improve the prediction of 
the dipole strength for the unstable nuclides
passed in the stellar events.
We chose the combination, IBA-1 and RPA with dipole-dipole
interaction, because it is simple and the modules were available. 
Our simple RPA can be replaced by
more sophisticated versions based on modern density functionals, which
are expected to provide reliable dipole strength functions for
unstable nuclei. Likewise, the IBA-1
phenomenology can be extended in a systematic way or it
can be replaced by any large-amplitude
description for the collective shape degrees of freedom.     
%The development of predictive models for the low-lying collective
%excitations in exotic nuclei is a central activity at present. 

In summary, we propose a novel method (ISS-RPA) for calculating
the dipole strength function of nuclei with large-amplitude
shape fluctuations, which combines the Interacting Boson Model (IBA)
with the Random Phase Approximation (RPA). 
The method bases on the existence of  two time scales:  the slow
shape dynamics and the fast dipole vibrations.
Instantaneous Shape Sampling (ISS) assumes that the 
photo-absorption occurs at a fixed shape, described by  RPA,
the probability of which given by IBA.
We studied the nuclides with ($Z$=38, 40, 42, $N$=50) and 
($Z$=42, $N$=52, 54, 56, 58). ISS-RPA very well reproduces 
the experimental photo-absorption
cross sections $\sigma_\gamma$.  Around the nucleon emission thresholds,
$\sigma_\gamma$ is determined by the Landau fragmentation and the fluctuating 
shapes. ISS-RPA may be used for calculating $\sigma_\gamma$ of  unstable
nuclei needed in  astrophysical simulations of violent stellar events  for
a firm understanding of the origin of chemical elements in the universe.

This work was supported by the DFG project KA2519/1-1 (Germany) and the
US DOE grant DE-FG02-95ER4093.

\end{document}